\documentclass[preprint,showpacs,preprintnumbers,amsmath,amssymb,showkeys]{revtex4}

\usepackage{dcolumn}
\usepackage{bm}
\usepackage{amssymb,amsmath}

\begin{document}


\title{Dual DSR}

\author{Jose A. Magpantay}
\email{jose.magpantay@up.edu.ph} \affiliation{National Institute
of Physics and Technology Management Center, University of the Philippines, Quezon City,
Philippines\\}

\date{\today}

\begin{abstract}
We develop the physics of dual kappa Poincare algebra, which we will call dual DSR. First we show that the dual kappa Poincare algebra is isomorphic to de Sitter algebra and its space-time is essentially de Sitter space-time. Second, we show how to derive the coproduct rule for Beltrami and conformal coordinates of de Sitter space-time. It follows from the current literature on de Sitter relativity that the speed of light c and the de Sitter length are the two invariant scales of the physics of dual kappa Poincare. Third, we derive the first Casimir invariant of the dual kappa Poincare algebra and use this to derive an expression for the speed of light, our fourth result. Fifth, the field equation for the scalar field is derived from the Casimir invariant. The results for the coordinate speed of light and the scalar field theory are the same as in de Sitter theory in the planar coordinate basis. Thus, we have shown that the physics of the dual kappa Poincare algebra (in the dual bicrossproduct basis), which can be appropriately called dual DSR, is essentially de Sitter relativity. Sixth, we argue the existence of an observer-independent minimum momentum. Seventh, we argue heuristically that the existence of minimum momentum will lead to a dual generalized uncertainty principle. Finally, we note that dual DSR is not a quantum theory of space-time but a quantum theory of momenta.
\end{abstract}

\maketitle

\section{\label{sec:level1}Introduction}
	Doubly special relativity (DSR), the physics of kappa Poincare algebra, is a bottom-up approach to quantum gravity and was initiated in \cite{Amelino-Camelia}. It leads to a quantum space-time at distance scales of the Planck length, which unfortunately seems to be contradicted by the recent measurement of the speed of light from a gamma ray burst \cite{Abdo}. But before this major problem came to light, DSR already faced some difficulties including the so called soccer ball problem (see \cite{Kowalski-Glikman} for an introductory discussion), which limits the energy of macroscopic objects to that of the Planck energy, something that is obviously false. The origin of this problem is the algebraic properties of space-time and momentum space, i.e., space-time has trivial co-algebra and non-trivial algebra while momentum space has non-trivial co-algebra and trivial algebra. The non-trivial co-algebra of the momentum space leads to the soccer ball problem while the non-trivial algebra of space-time leads to a quantum gravity that linearly alters the speed of light with energy. As already noted, the former is obviously false while the latter seems to be particularly ruled out by the recent measurement of the speed of light from gamma ray burst \cite{Abdo}.

	When the author presented the dual kappa Poincare algebra \cite{Magpantay}, one motivation was to avoid the soccer ball problem. An obvious way to do this is to have a trivial co-algebra in momentum space. Following the chain of arguments in kappa Poincare algebra and its Hopf algebra, the author constructed the dual kappa Poincare algebra with the following properties - (1) the Lorentz subalgebra remains the same, (2) the space-time has a trivial algebra but non-trivial co-algebra, (3) the momentum space has trivial co-algebra but non-trivial algebra, (4) the space-time is not unique as it satisfies a relation similar to the equation satisfied by the momenta in kappa Poincare algebra.

	In this paper, we will develop the physics of dual kappa Poincare algebra. First, we will show that the dual kappa Poincare algebra is isomorphic to de Sitter algebra. Then we will complete the proof that the dual kappa Poincare space-time is the same as de Sitter space-time by showing that the different coordinate systems used in de Sitter relativity satisfy the general relation satisfied by dual kappa Poincare space-time.  These two suggest that the physics of dual kappa Poincare algebra is the same as de Sitter relativity, which is known to have two invariant scales - the speed of light c and the de Sitter length \cite{Aldrovandi} \cite{Guo}.

	As a byproduct of the equivalence of dual kappa Poincare space-time to de Sitter space-time, we present the coproducts of space-time in the natural, Beltrami and conformal coordinates starting from the coproduct rule in the planar coordinate basis. The coproducts in the Beltrami and conformal coordinates are systematically expanded in powers of $\bar{\kappa}$, which is an excellent expansion parameter because of its extremely small value (given the present size of the universe, about $10^{-26} m^{-1})$ \cite{Guo}.

	Next we derive the first Casimir invariant of dual kappa Poincare algebra. Unlike in kappa Poincare algebra where the first Casimir invariant is purely a function of momenta, the first Casimir of the dual kappa Poincare algebra depends on all the generators - the momenta, boosts and rotations. From the first Casimir invariant, we derive an expression for the speed of light. The speed of light we get is the coordinate speed of light in the planar coordinate system of de Sitter space-time.

	We next derive the field equation for a scalar particle from the first Casimir invariant. Since space-time commutes, the scalar field theory derivation is rather straightforward unlike in kappa Poincare algebra that involved a five-dimensional differential calculus. In the case of dual kappa Poincare, we only needed to find a differential realization of the momenta that is consistent with the algebra. The scalar field theory is that of a scalar field theory in de Sitter gravity. 
	
	If in DSR there is a minimum length - the Planck length - that is observer-independent, we next argue that there must be a minimum momentum in dual DSR or de Sitter relativity. This $P_{min}$, from the de Broglie relation must be equal to $h/\textit{l}$, where $\textit{l}$ is the de Sitter length. We then argue that this minimum momentum is also observer-independent. Furthermore, we derive, heuristically, that the existence of this minimum momentum changes the uncertainty principle to what can be called dual generalized uncertainty principle.   
	
	All these mean that dual DSR is de Sitter relativity, where space-time is classical (commutes) and the momenta is quantized (non-commuting). The ideas presented here have intersections with the ideas of Aldrovandi et al \cite{Aldrovandi} and Guo et al \cite{Guo}. Guo et al argued explicitly that de Sitter relativity in Beltrami coordinates is dual to Snyder's theory of non-commuting spacetime (which is one of the basis for DSR theories) and even pointed out the dual relation between the Planck length (invariant length in DSR) and the de Sitter length (invariant length in de Sitter). In another paper, Aldrovandi et al \cite{Aldrovandi2}, showed how the Heisenberg algebra is modified in de Sitter space-time. The Heisenberg algebra they derived is dual to the Heisenberg algebra of quantum gravity that arise at the Planck length.  
	
	To put this work in perspective, Bacry and Le Blond \cite{Bacry} first presented the de Sitter group when they argued kinematically the possible Lie algebras satisfied by translations, real rotations and Lorentz boosts under the assumptions of spacetime isotropicity and parity and time-reversal automorphism. Contractions to various groups, including Poincare and Newton-Hooke groups, were also first discussed in this paper. Since then, many papers, including those by the groups cited above and those cited in the next section looked into the various aspects of de Sitter spacetime. A comprehensive discussion of de Sitter relativity is found in the recent paper of Cacciatori et al \cite{Cacciatori}. The present author's derivation of de Sitter algebra made use of the coalgebra and the Heisenberg double construction used in kappa Poincare deformation thus the label dual kappa Poincare algebra. The direct connection with de Sitter algebra is done in this paper by proving that the two are isomorphic. Also in this work, we will show that the coordinate systems used in de Sitter spacetime satisfy the general condition for dual kappa Poincare spacetime, thereby hinting that the physics of dual kappa Poincare algebra is de Sitter relativity. The results on the Casimir invariant of dual kappa Poincare, from which we derive the speed of light and scalar field theory are all known in de Sitter relativity except it was presented here in the way how things are derived in DSR. The new contribution in this part is the derivation of the coproduct rules in the various coordinate systems used in de Sitter space-time. The observer-independent minimum momentum, which leads to a dual generalized uncertainty principle are the new contributions in the field of de Sitter relativity.  
\section{\label{sec:level2}Equivalence of Dual Kappa Poincare and de Sitter}
	In a recent paper \cite{Magpantay}, the author derived another way of modifying the Poincare algebra such that the Lorentz algebra is maintained, space-time commutes (but does not co-commute) while the momentum space co-commutes (but does not commute). By structure and construction this algebra is dual to the kappa Poincare algebra in the bicrossproduct basis and thus appropriately called dual kappa Poincare algebra. The algebra is given by the following relations:
\begin{subequations}\label{sha}
\begin{gather}
\left[\bar{M_i},\bar{M_j}\right]=i\epsilon_{ijk}\bar{M_k},\label{first}\\
\left[\bar{N_i},\bar{N_j}\right]=-i\epsilon_{ijk}\bar{M_k},\label{second}\\
\left[\bar{M_i},\bar{N_j}\right]=i\epsilon_{ijk}\bar{N_k},\label{third}\\
\left[\bar{M_i},\bar{P_0}\right]=0,\label{fourth}\\
\left[\bar{M_i},\bar{P_j}\right]=i\epsilon_{ijk}\bar{P_k},\label{fifth}\\
\left[\bar{N_i},\bar{P_0}\right]=i\bar{P_i}+i\bar\kappa\bar{N_i},\label{sixth}\\
\left[\bar{N_i},\bar{P_j}\right]=i\delta_{ij}\bar{P_0}+i\bar\kappa\epsilon_{ijk}\bar{M_k},
\end{gather}
\end{subequations}
where $\bar{M_i}$, $\bar{N_i}$ and $\bar{P_\mu}$ are the rotation, boost and momentum generators. Note, we corrected the sign errors in equations (1b) and (1g) that was given in \cite{Magpantay}. The parameter $\bar\kappa$ is related to the cosmological constant $\Lambda$ and de Sitter length $\textit{l}$ via
\begin{equation}\label{sha2}
\bar\kappa=\Lambda^{\frac{1}{2}}=\textit{l}^{-1},
\end{equation}

We will now show that this algebra is isomorphic to de Sitter algebra.

	The de Sitter algebra is geometrically understood by embedding de Sitter space-time in a five-dimensional Minkowski space-time through the hyperboloid \cite{Aldrovandi2}, \cite{Mignemi}, \cite{Spradlin}.
\begin{equation}\label{sha3}
\eta^{AB}y_Ay_B=y_0^2-\vec{y}\cdot\vec{y}-y_4^2=-l^2,
\end{equation}
where $\textit{l}$ is the de Sitter length and $\eta^{AB}=(+,-,-,-,-)$ is the five-dimensional metric and capital Latin letters run from 0 to 4. The isometries of the hyperboloid are generated by the so(1,4) algebra, the Lorentz algebra of the five-dimensional Minkowski space, which are also the generators of the de Sitter algebra. The generators are given by
\begin{equation}\label{sha4}
J_{AB}=y_A\pi_B-y_B\pi_A,
\end{equation}
where $\pi_A=i\eta_{AB}\frac{\partial}{\partial{y_B}}$. The so(1,4) algebra is given by
\begin{equation}\label{sha5}
\left[J_{AB},J_{CD}\right]=i\eta_{BC}J_{AD}+i\eta_{AD}J_{BC}-i\eta_{AC}J_{BD}-i\eta_{BD}J_{AC}.
\end{equation}
This equation contains the so(1,3) Lorentz algebra of the 4 dimensional Minkowski space-time, which is the same as equation (5) except the indices A, B etc. are replaced by lower case Greek letters $\mu$, $\nu$, etc. that run from 0 to 3. The so(1,3) algebra can be rewritten as
\begin{subequations}\label{sha6}
\begin{gather}
\left[m_i,m_j\right]=i\epsilon_{ijk}m_k,\label{first}\\
\left[m_i,n_j\right]=i\epsilon_{ijk}n_k,\label{second}\\
\left[n_i,n_j\right]=-i\epsilon_{ijk}m_k,
\end{gather}
\end{subequations}
where the indices (i,j,k) run from 1 to 3 and
\begin{subequations}\label{sha7}
\begin{gather}
m_i=\epsilon_{ijk}J_{jk},\label{first}\\
n_i=J_{0i}.
\end{gather}
\end{subequations}
From equations (5) and (6) we can identify the the Lorentz subalgebra of the 4D de Sitter algebra with the flat space Lorentz algebra, i.e.,
\begin{subequations}\label{sha8}
\begin{gather}
m_i=\epsilon_{ijk}x_jp_k,\label{first}\\
n_i=x_0p_i-x_ip_0,
\end{gather}
\end{subequations}
where $(x_\mu,p_\mu)$ are the Minkowski phase space variables.
	The realization of the translation generators of de Sitter space depend on the choice of coordinates on the hyperboloid. However, the translation generator is defined \cite{Mignemi}
\begin{equation}\label{sha9}
\tilde{P_\mu}=\bar{\kappa}J_{4\mu}.
\end{equation}
Using equation (5), we find that the translation generators satisfy
\begin{subequations}\label{sha10}
\begin{gather}
\left[J_{\mu\nu},\tilde{P_\rho}\right]=\eta_{\mu\rho}\tilde{P_\nu}-\eta_{\nu\rho}\tilde{P_\mu},\label{first}\\
\left[\tilde{P_\mu},\tilde{P_\nu}\right]=-\bar{\kappa}^2J_{\mu\nu}.
\end{gather}
\end{subequations}
Equations (5) (with A,B replaced by $\mu$, $\nu$), (10a) and (10b) constitute the de Sitter algebra. The accompanying phase space algebra and the action of rotations and boosts on space-time depend on the coordinate system used.
	
	Equations (1a) to (1c) of the dual kappa Poincare algebra are the same as equations (6a) to (6c), they represent the Lorentz subalgebra. Equations (1d) and (1e) are the same as equations (10a) with $\mu=i, \nu=j$. However, equations (1f) and (1g) are not the same as equation (10a) with $\mu=i, \nu=0$. Lastly, in dual kappa Poincare algebra
\begin{subequations}\label{sha11}
\begin{gather}
\left[\bar{P_0},\bar{P_i}\right]=i\bar{\kappa}\bar{P_i},\label{first}\\
\left[\bar{P_i},\bar{P_j}\right]=0,
\end{gather}
\end{subequations}
which are definitely not the same as equation (10b) of de Sitter algebra. Thus, it looks like the dual kappa Poincare algebra is not the same as de Sitter algebra.
	However, consider the following simple transformations:
\begin{subequations}\label{sha12}
\begin{gather}
\tilde{N_i}=\bar{N_i},\label{first}\\
\tilde{M_i}=\bar{M_i},\label{second}\\
\tilde{P_0}=\bar{P_0},\label{third}\\
\tilde{P_i}=\bar{P_i}+\bar{\kappa}\bar{N_i}.
\end{gather}
\end{subequations}
Equations (12a) and (12b) guarantee the same Lorentz subalgebra. Equations (12b) and (12c) will also leave unchanged equations (1d) and (1e), which are the same as in de Sitter algebra. Using equations (1f) and (1g) and equations (12a) and (12b), we find
\begin{subequations}\label{sha13}
\begin{gather}
\left[\tilde{N_i},\tilde{P_j}\right]=i\delta_{ij}\tilde{P_0},\label{first}\\
\left[\tilde{N_i},\tilde{P_0}\right]=i\tilde{P_i},\label{second}\\
\left[\tilde{P_0},\tilde{P_i}\right]=-i\bar{\kappa}^2\tilde{N_i},\label{third}\\
\left[\tilde{P_i},\tilde{P_j}\right]=-i\bar{\kappa}^2\epsilon_{ijk}\tilde{M_k}.
\end{gather}
\end{subequations}
Now, equations (13a) and (13b) are the same as equation (10a) for $\mu=0$, $\nu=i$ while equations (13c) and (13d) are the same as equation (10b). Thus, by a simple transformation defined by equation (12d) the dual kappa Poincare algebra is isomorphic to de Sitter algebra.
	
	Since dual kappa Poincare algebra is isomorphic to de Sitter algebra, their space-times must be the same. And indeed this will be shown true to be true in the following. Reference (3) clearly shows that the space-time of kappa Poincare is not unique. Expressed in terms of a reference Minkowski space-time, the dual kappa Poincare space-time are given by
\begin{subequations}\label{sha14}
\begin{gather}
\bar{X_0}=\bar{f}(x_0,\vec{x}\cdot\vec{x}),\label{first}\\
\bar{X_i}=x_i\bar{g}(x_0,\vec{x}\cdot\vec{x}),
\end{gather}
\end{subequations}
with inverses given by
\begin{subequations}\label{sha15}
\begin{gather}
x_0=\bar{F}(\bar{X_0},\vec{\bar{X}}\cdot\vec{\bar{X}}),\label{first}\\
x_i=\bar{X_i}\bar{G}(\bar{X_0},\vec{\bar{X}}\cdot\vec{\bar{X}}).
\end{gather}
\end{subequations}
Equations (14a) and (14b) are are derived by imposing that the rotation generators $\bar{M_i}$ leave time invariant and rotate the spatial coordinates. The action of boosts $\bar{N_i}$ on space-time are given by
\begin{subequations}\label{sha16}
\begin{gather}
\left[\bar{N_i},\bar{X_0}\right]=i\bar{D}(\bar{X_0},\vec{\bar{X}}\cdot\vec{\bar{X}})\bar{X_i},\label{first}\\
\left[\bar{N_i},\bar{X_j}\right]=i\delta_{ij}\bar{A}(\bar{X_0},\vec{\bar{X}}\cdot\vec{\bar{X}})+i\bar{X_i}\bar{X_j}\bar{B}(\bar{X_0},\vec{\bar{X}}\cdot\vec{\bar{X}}),
\end{gather}
\end{subequations}
where
\begin{subequations}\label{sha17}
\begin{gather}
\bar{A}=\bar{F}\bar{g},\label{first}\\
\bar{B}=\bar{G}^2\left(2\bar{F}\frac{\partial\bar{g}}{\partial(\vec{x}\cdot\vec{x})}+\frac{\partial\bar{g}}{\partial x_0}\right),\label{second}\\
\bar{D}=\bar{G}\left(2\bar{F}\frac{\partial\bar{f}}{\partial(\vec{x}\cdot\vec{x})}+\frac{\partial\bar{f}}{\partial x_0}\right).
\end{gather}
\end{subequations}
Since $\bar{A}$, $\bar{B}$ and $\bar{D}$ satisfy the non-linear equation
\begin{equation}\label{sha18}
\frac{\partial{\bar{A}}}{\partial{\bar{X_0}}}\bar{D}+2\frac{\partial{\bar{A}}}{\partial{(\vec{\bar{X}}\cdot\vec{\bar{X}})}}\left[\bar{A}+(\vec{\bar{X}}\cdot\vec{\bar{X}})\bar{B}\right]-\bar{A}\bar{B}=1,
\end{equation}
out of the two functions $\bar{f}$, $\bar{g}$ (or equivalently $\bar{F}$, $\bar{G}$), one is left unspecified. Thus, there is no unique dual kappa Poincare space-time. Equations (14) to (18) are similar to the relations satisfied by the momenta in kappa Poincare algebra, thus the label dual kappa Poincare algebra is appropriate for the algebra we are considering here.

	We will now show that the known coordinate systems used in describing de Sitter space-time satisfy equation (18). In reference (3), the dual kappa Poincare algebra in the dual bicrossproduct basis was presented. Its space-time $\bar{X_\mu}$ is given in terms of a reference Minkowski space-time $x_\mu$ by the transformations
\begin{subequations}\label{sha19}
\begin{gather}
\bar{X_0}=\bar{f}=\frac{1}{\bar{\kappa}}\ln{(\bar{\kappa}x_0+\sqrt{\bar{\kappa}^2(x_0^2-\vec{x}\cdot\vec{x})+1})},\label{first}\\
\bar{X_i}=x_i\bar{g}=x_i\dfrac{1}{\bar{\kappa}x_0+\sqrt{\bar{\kappa}^2(x_0^2-\vec{x}\cdot\vec{x})+1}},
\end{gather}
\end{subequations}
with the inverses
\begin{subequations}\label{sha20}
\begin{gather}
x_0=\bar{F}=\frac{1}{\kappa}\sinh{(\bar{\kappa}\bar{X_0})}+\frac{\bar{\kappa}}{2}\vec{\bar{X}}\cdot\vec{\bar{X}}\exp{(\bar{\kappa}\bar{X_0})},\label{first}\\
x_i=\bar{X_i}\bar{G}=\bar{X_i}\exp{(\bar{\kappa}\bar{X_0})}.
\end{gather}
\end{subequations}
Equations (19) and (20) define the planar coordinates of de Sitter space-time \cite{Spradlin}, which is usually presented with the fifth coordinate of the hyperboloid given by
\begin{equation}\label{sha21}
x_4=\frac{1}{\bar{\kappa}}\cosh{\frac{\bar{X_0}}{\bar{\kappa}}}-\frac{\bar{\kappa}}{2}(\vec{\bar{X}}\cdot \vec{\bar{X}}) \exp{(\bar{\kappa}\bar{X_0})}.
\end{equation}
What is the relevance of completing the Minkowski space-time $(x_0,\vec{x})$ to de Sitter space-time $(x_0,\vec{x},x_4)$ in defining the dual kappa Poincare space-time $(\bar{X_0},\vec{\bar{X}})$? Kowalski-Glikman \cite{Kowalski-Glikman2} showed that the geometrical basis of the bicrossproduct kappa Poincare algebra is de Sitter momentum space. Since the dual bicrossproduct basis of the dual kappa Poincare algebra has the same structure as the bicrossproduct basis of kappa Poincare algebra in momentum space for the latter but in space-time for the former, then the same argument for the completion to de Sitter space-time holds. In particular, without the completion to de Sitter space-time equation (19) would have led to a trivial co-product instead of
\begin{subequations}\label{sha22}
\begin{gather}
\Delta(\bar{X_0})=\bar{X_0}\otimes\textbf{1}+\textbf{1}\otimes\bar{X_0},\label{first}\\
\Delta(\bar{X_i})=\bar{X_i}\otimes\textbf{1}+\exp{(-\bar{\kappa}\bar{X_0})}\otimes\bar{X_i}.
\end{gather}
\end{subequations}
	
	By construction of the dual kappa Poincare algebra, its space-time is not unique because one of the two functions $\bar{f}$ and $\bar{g}$ (or equivalently $\bar{F}$ and $\bar{G}$) is left unspecified. The dual kappa Poincare space-time satisfies a non-linear constraint given by equations (17) and (18). In the case of the dual bicrossproduct basis, which is the same as the planar coordinates of de Sitter space-time, equations (17), (19) and (20) yield
\begin{subequations}\label{sha23}
\begin{gather}
\bar{A_p}=\frac{1}{2\bar{\kappa}}(1-\exp{(-2\bar{\kappa}\bar{X_0})})+\frac{\bar{\kappa}}{2}(\vec{\bar{X}}\cdot\vec{\bar{X}}),\label{first}\\
\bar{B_p}=-\bar{\kappa},\label{second}\\
\bar{D_p}=1.
\end{gather}
\end{subequations}
These satisfy equation (18).
	What we will show is that some of the other known coordinates used in de Sitter space-time also satisfy equation (18) and this should conclusively prove that the space-time of dual kappa Poincare is de Sitter space-time. We have not verified that all the coordinate systems used in describing de Sitter space satisfy equation (18) but since these different coordinate systems are different ways of parametrizing the hyperboloid given by equation (3), then the different coordinate systems can be related to one another. And if one satisfies equation (18), the others should also satisfy equation (18). Thus we will only verify the dual kappa Poincare space-time relation for a few of de Sitter space-time coordinate systems.

	(a)	Natural coordinates

This is the simplest case since
\begin{equation}\label{sha24}
y_\mu=X^n_\mu.
\end{equation}
The fifth coordinate of the hyperboloid is given by
\begin{equation}\label{25}
y^n_4=\sqrt{(X^n)^2+l^2}.
\end{equation}
From equation (24) we read off
\begin{subequations}\label{sha26}
\begin{gather}
\bar{f_n}=y_0,\label{first}\\
\bar{g_n}=1,\label{second}\\
\bar{F_n}=X^n_0,\label{third}\\
\bar{G_n}=1.
\end{gather}
\end{subequations}
from which we get
\begin{subequations}\label{sha27}
\begin{gather}
\bar{A_c}=X^c_0,\label{first}\\
\bar{B_c}=0,\label{second}\\
\bar{D_c}=1,
\end{gather}
\end{subequations}
which obviously satisfy equation(18). Thus, we have shown that the natural coordinate system used in de Sitter space-time satisfy the dual kappa Poincare spacetime condition.
	
	The coproduct rule for the natural coordinates of de Sitter spacetime (equivalent to equation (22)) easily follows from the classical basis of DSR \cite{Kowalski-Glikman3} because the natural coordinates of de Sitter spacetime is dual to the classical basis of DSR. From the coproduct rule of the momenta in the classical basis of DSR, we infer the coproduct rule for the natural coordinates given by
\begin{subequations}\label{sha28}
\begin{gather}
\Delta(X^n_i)=X^n_i\otimes\textbf{K}+\textbf{1}\otimes X^n_i,\label{first}\\
\Delta(X^n_0)=\frac{1}{2\bar{\kappa}}\left(\textbf{K}\otimes\textbf{K}-\textbf{K}^{-1}\otimes\textbf{K}^{-1}\right)+\textbf{H}
\end{gather}
\end{subequations}
where
\begin{subequations}\label{sha29}
\begin{gather}
\textbf{K}=\bar{\kappa}\left[X^n_0+((X^n_0)^2-\vec{X^n}\cdot\vec{X^n}+l^2)^{\frac{1}{2}}\right],\label{first}\\
\textbf{H}=\frac{\bar{\kappa}}{2}\left(\textbf{K}^{-1}\vec{X^n}\cdot\vec{X^n}\otimes\textbf{K}+2\textbf{K}^{-1}X^n_i\otimes X^n_i+\textbf{K}^{-1}\otimes\textbf{K}^{-1}\vec{X^n}\cdot\vec{X^n}\right)
\end{gather}
\end{subequations}

		(b)	Beltrami coordinates

The Beltrami coordinates \cite{Mallett} of de Sitter space-time are given by
\begin{equation}\label{sha30}
y_\mu=\dfrac{X^b_\mu}{\sqrt{1-\bar{\kappa}^2((X^b_0)^2-\vec{X^b}\cdot\vec{X^b})}}.
\end{equation}
The fifth coordinate of the hyperboloid in terms of the Beltrami coordinates is
\begin{equation}\label{sha31}
y^b_4=(1-\bar{\kappa}^2X^2)^{-\frac{1}{2}}.
\end{equation}
The inverse of equation(30) gives
\begin{equation}\label{sha32}
X_\mu=\frac{y_\mu}{\sqrt{1+\bar{k}^2y^2}}.
\end{equation}
From these we derive,
\begin{subequations}\label{sha33}
\begin{gather}
\bar{f_b}=\dfrac{y_0}{\sqrt{1+\bar{\kappa}^2y^2}},\label{first}\\
\bar{g_b}=\frac{1}{\sqrt{1+\bar{\kappa}^2y^2}},\label{second}\\
\bar{F_b}=\dfrac{X_0}{\sqrt{1-\bar{\kappa}^2X^2}},\label{third}\\
\bar{G_b}=\frac{1}{\sqrt{1-\bar{\kappa}^2X^2}}.
\end{gather}
\end{subequations}
From equations (33a) to (33d), we get $\bar{A_b}$, $\bar{B_b}$ and $\bar{D_b}$ with values the same as those given by equation (27), which means the Beltrami coordinates of de Sitter space also satisfies the condition for space-time of dual kappa Poincare.

	We will now derive the coproduct rule for Beltrami coordinates. We can easily relate the Beltrami coordinates to the planar coordinates by making use of the relations of both coordinate systems to the hyperboloid. The result is
\begin{subequations}\label{sha34}
\begin{gather}
\bar{X}_0^B=\dfrac{\frac{1}{\kappa}\sinh{(\bar{\kappa}\bar{X}_0)}+\frac{\bar\bar{\kappa}}{2}(\vec{\bar{X}}\cdot\vec{\bar{X}})\exp{(\bar{\kappa}\bar{X}_0)}}{(1+\Gamma)^{\frac{1}{2}}},\label{first}\\
\bar{X}_i^B=\dfrac{\bar{X}_i\exp{(\bar{\kappa}\bar{X}_0)}}{(1+\Gamma)^{\frac{1}{2}}},\label{second}\\
\Gamma=\sinh^2{(\bar{\kappa}\bar{X}_0)}+\frac{\bar{\kappa}^4}{4}\left(\vec{\bar{X}}\cdot\vec{\bar{X}}\right)^2\exp{(2\bar{\kappa}\bar{X}_0)} -\frac{1}{2}\bar{\kappa}^2\left(\vec{\bar{X}}\cdot\vec{\bar{X}}\right)\exp{(2\bar{\kappa}\bar{X}_0)}-\frac{1}{2}\bar{\kappa}^2\left(\vec{\bar{X}}\cdot\vec{\bar{X}}\right).
\end{gather}
\end{subequations}
The coproduct rules can be systematically expanded in powers of $\bar{\kappa}$, which is a valid expansion parameter because its value is very small. We do this by (i) doing a Maclaurin expansion of equations (34a) to (34c) in terms of $\bar{X}_0$ and $\vec{\bar{X}}\cdot\vec{\bar{X}}$ (note, there is no operator ordering problem because the coordinates commute), (ii) using the coproduct rules for the planar coordinates given by equation (22), (iii) employing the following coproduct rules
\begin{subequations}\label{sha35}
\begin{gather}
\Delta(AB)=\Delta(A)\Delta(B),\label{first}\\
(A\otimes B)(C\otimes D)=AC\otimes BD,
\end{gather}
\end{subequations}
and (iv) expressing the planar coordinates in terms of the Beltrami coordinates as given by
\begin{subequations}\label{sha36}
\begin{gather}
\bar{X}_i=\dfrac{\bar{X}_i^B}{1+\bar{\kappa}\bar{X}_0^B},\label{first}\\
\bar{X}_0=\frac{1}{\bar{\kappa}}\ln{\left[\dfrac{1+\bar{\kappa}\bar{X_0^B}}{\left(1-\bar{\kappa}^2((\bar{X^B_0})^2-\vec{\bar{X^B}}\cdot\vec{\bar{X^B}})\right)^{\frac{1}{2}}}\right]}.
\end{gather}
\end{subequations}
The resulting coproduct rule for the Beltrami coordinates, expanded up to order $\bar{\kappa}$, are
\begin{subequations}\label{sha37}
\begin{gather}
\Delta(\bar{X}_i^B)=\bar{X_i^B}\otimes\textbf{1}+\textbf{1}\otimes\bar{X_i^B}+\bar{\kappa}[\bar{X_i^B}\otimes\bar{X_0^B}+\textbf{1}\otimes\bar{X_i^B}\bar{X_0^B}]+ \ldots,\label{first}\\
\begin{split}
\Delta(\bar{X}_0^B)&=\bar{X}_0^B\otimes\textbf{1}+\textbf{1}\otimes\bar{X}_0^B+\frac{\bar{\kappa}}{2}[(\vec{\bar{X^B}}\cdot\vec{\bar{X^B}})\otimes\textbf{1}+\textbf{1}\otimes(\vec{\bar{X^B}}\cdot\vec{\bar{X^B}})\\&-(\bar{X}_0^B)^2\otimes\textbf{1}-\textbf{1}\otimes(\bar{X}_0^B)^2+2\bar{X}_i^B\otimes\bar{X}_i^B]+ \ldots.
\end{split}
\end{gather}
\end{subequations}
		
	(c)	Conformal coordinates

	This is arrived at through stereographic projections from the hyperboloid coordinates $y_A$ (see equation 3) to a 4D space-time $(X^c_0,X^c_i)$ and it is given by \cite{Aldrovandi}, \cite{Gursey}.
\begin{equation}\label{sha38}
y_\mu=\dfrac{1}{[1-\frac{1}{4l^2}((X^c_0)^2-\vec{X^c}\cdot\vec{X^c})]}X^c_\mu.
\end{equation}
For completeness and comparison with the planar coordinates, we give the fifth component of the hyperboloid in conformal coordinates
\begin{equation}\label{sha39}
y^c_4=\sqrt{l^2+y^2}=l\dfrac{(1+\frac{X^2}{4l^2})}{(1-\frac{X^2}{4l^2})},
\end{equation}
where $y^2=y_0^2-\vec{y}\cdot\vec{y}$. The inverse is given by
\begin{equation}\label{sha40}
\begin{split}
X^c_\mu&=-\frac{2l^2}{y^2}\left(1\pm\sqrt{1+\frac{y^2}{l^2}}\right)y_\mu\\
&=\dfrac{2y_\mu}{1+(1+\bar{\kappa}^2y^2)}
\end{split}
\end{equation}
From these we identify
\begin{subequations}\label{sha41}
\begin{gather}
\bar{f_c}=-\frac{2l^2}{y^2}\left(1\pm\sqrt{1+\frac{y^2}{l^2}}\right)y_0,\label{first}\\
\bar{g_c}=-\frac{2l^2}{y^2}\left(1\pm\sqrt{1+\frac{y^2}{l^2}}\right),\label{second}\\
\bar{F_c}=\left(1-\frac{(X^c)^2}{4l^2}\right)^{-1}X^c_0,\label{third}\\
\bar{G_c}=\left(1-\frac{(X^c)^2}{4l^2}\right)^{-1}.
\end{gather}
\end{subequations}
If the conformal coordinates of de Sitter is to satisfy the condition for dual kappa Poincare space-time, the corresponding $\bar{A_c}$, $\bar{B_c}$ and $\bar{D_c}$ should satisfy equation (18). Indeed this is true since the values are the same as those given in equation (27).

	To derive the coproduct rules for the conformal coordinates, we follow the same procedure we used in the Beltrami coordinates. Using the conformal coordinates - hyperboloid coordinates relation given by equation (40) (we will use the second relation) and the hyperboloid - planar coordinates relation given by equation (15), we derive the conformal-planar coordinates relations
\begin{subequations}\label{sha42}
\begin{gather}
\bar{X}_0^c=\dfrac{\frac{2}{\bar{\kappa}}\sinh{(\bar{\kappa}\bar{X}_0)}+\bar{\kappa}(\vec{\bar{X}}\cdot\vec{\bar{X}})\exp{\bar{\kappa}\bar{X}_0}}{\left[1+(1+\Gamma)^{\frac{1}{2}}\right]},\label{first}\\
\bar{X}_i^c=\dfrac{2\bar{X}_i\exp{\bar{\kappa}\bar{X}_0}}{\left[1+(1+\Gamma)^{\frac{1}{2}}\right]},
\end{gather}
\end{subequations}
where $\Gamma$ is given by equation (34c). Doing a Maclaurin series expansion and making use of the coproduct rules given by equation (35) and making use of the inverse relations to equations (42a) and (42b), which gives
\begin{subequations}\label{sha43}
\begin{gather}
\bar{X}_0=\frac{1}{\bar{\kappa}}\ln{\left[\dfrac{1+\bar{\kappa}\bar{X}_0^c+\frac{\bar{\kappa}^2}{4}(\vec{\bar{X}^c}\cdot\vec{\bar{X}^c})}{1-\frac{\bar{\kappa}^2}{4}(\vec{\bar{X}^c}\cdot\vec{\bar{X}^c})}\right]},\label{first}\\
\bar{X}_i=\dfrac{\bar{X}_i^c}{1+\bar{\kappa}\bar{X}_0^c+\frac{\bar{\kappa}^2}{4}(\vec{\bar{X}^c}\cdot\vec{\bar{X}^c})},
\end{gather}
\end{subequations}
we get the coproduct rules up to order $\bar{\kappa}$
\begin{subequations}\label{sha44}
\begin{gather}
\Delta(\bar{X}_0^c)=\bar{X}_0^c\otimes\textbf{1}+\textbf{1}\otimes\bar{X}_0^c+\bar{\kappa}\bar{X}_i^c\otimes\bar{X}_i^c+\ldots,\label{first}\\
\Delta(\bar{X}_i^c)=\bar{X}_i^c\otimes\textbf{1}+\textbf{1}\otimes\bar{X}_i^c+\bar{\kappa}\bar{X}_i^c\otimes\bar{X}_0^c+\ldots.
\end{gather}
\end{subequations}

	Thus, we have established that the space-time of dual kappa Poincare algebra is de Sitter space-time. We also worked out the coproduct rules for some of the coordinate systems used in de Sitter space-time. The lowest order corrections to flat space-time coproducts are given by the $\bar{\kappa}$ terms in equations (37) and (43). From \cite{Aldrovandi}, \cite{Guo}, which showed that de Sitter relativity has two invariant scales - the speed of light and de Sitter length, it follows that the physics of dual kappa Poincare algebra can be appropriately called dual DSR.
\section{\label{sec:level3}The Casimir Invariant and the Speed of Light}
	The Casimir invariants of an algebra commute with all the generators of that algebra. The first Casimir invariant is particularly significant because the speed of light and the scalar field theory are derivable from it. In the case of Poincare algebra, where the first Casimir is given by
\begin{equation}\label{sha45}
C(p_0,\vec{p}\cdot\vec{p})=p_0^2-\vec{p}\cdot\vec{p}=m^2.
\end{equation}
The speed of light is given by
\begin{equation}\label{sha46}
\left(\frac{\partial{p_0}}{\partial{p}}\right)_{m\rightarrow0}=1.
\end{equation}
In the case of kappa Poincare algebra in the bicrossproduct basis with algebra given by
\begin{subequations}\label{sha47}
\begin{gather}
\left[M_i,M_j\right]=i\epsilon_{ijk}M_k,\label{first}\\
\left[M_i,N_j\right]=i\epsilon_{ijk}N_k,\label{second}\\
\left[N_i,N_j\right]=-i\epsilon_{ijk}M_k,\label{third}\\
\left[M_i,P_0\right]=0,\label{fourth}\\
\left[M_i,P_j\right]=i\epsilon_{ijk}P_k,\label{fifth}\\
\left[N_i,P_0\right]=iP_i,\label{sixth}\\
\left[N_i,P_j\right]=i\delta_{ij}\left[\frac{\kappa}{2}\left(1-\exp{(-2\frac{P_0}{\kappa})}\right)+\frac{1}{2\kappa}\vec{P}\cdot\vec{P}\right]-\frac{i}{\kappa}P_iP_j,
\end{gather}
\end{subequations}
the first Casimir invariant is given by
\begin{equation}\label{sha48}
\textbf{C}(P_0,\vec{P}\cdot\vec{P})=\left[2\kappa\sinh(\frac{P_0}{2\kappa})\right]^2-(\vec{P}\cdot\vec{P})\exp{(\frac{P_0}{\kappa})}=m^2.
\end{equation}
The second equality in equation (48) follows from the Poincare limit ($\kappa\rightarrow\infty$), which gives $P_0\rightarrow p_0$ and $P_i\rightarrow p_i$. From this Casimir invariant, we derive the speed of light
\begin{equation}\label{sha49}
\left(\frac{\partial{P_0}}{\partial{P}}\right)_{m\rightarrow0}\approx(1+\frac{P}{\kappa}).
\end{equation}
Equation (49) gives an energy or frequency dependent speed of light that apparently is not consistent with the recent gamma ray burst experiment \cite{Abdo}.

	We will follow the same procedure in deriving the speed of light in dual kappa Poincare algebra. However, in this case, we find that the Casimir invariant cannot just be a function of $\bar{P_0}$ and $\vec{\bar{P}}\cdot\vec{\bar{P}}$. The dual kappa Poincare algebra is given by equation (1) while its phase space algebra is given by
\begin{subequations}\label{sha50}
\begin{gather}
\left[\bar{X_\mu},\bar{X_\nu}\right]=0,\label{first}\\
\left[\bar{P_i},\bar{X_0}\right]=\left[\bar{P_i},\bar{P_j}\right]=0,\label{second}\\
\left[\bar{P_0},\bar{X_0}\right]=i,\label{third}\\
\left[\bar{P_i},\bar{X_j}\right]=-i\delta_{ij},\label{fourth}\\
\left[\bar{P_0},\bar{X_i}\right]=-i\bar{\kappa}\bar{X_i},\label{fifth}\\
\left[\bar{P_0},\bar{P_i}\right]=i\bar{\kappa}\bar{P_i}.
\end{gather}
\end{subequations}
These equations mean that in dual kappa Poincare, space-time commute but not co-commute and the momenta co-commute but do not commute, which are opposite to those in kappa Poincare \cite{Magpantay}. If we look for a Casimir invariant that is purely a function of $\bar{P_0}$ and $\vec{\bar{P}}\cdot\vec{\bar{P}}$, the algebra given by equation (1) will yield $\bar{\textbf{C}}=0$, i.e., a trivial Casimir invariant. Thus, we need to expand the dependence of the Casimir. Since the Casimir invariants are scalars, it should depend only on $\vec{\bar{M}}\cdot\vec{\bar{M}}$, $\vec{\bar{N}}\cdot\vec{\bar{N}}$, $\vec{\bar{M}}\cdot\vec{\bar{N}}+\vec{\bar{N}}\cdot\vec{\bar{M}}$ and $\vec{\bar{N}}\cdot\vec{\bar{P}}+\vec{\bar{P}}\cdot\vec{\bar{N}}$. There is no $\vec{\bar{M}}\cdot\vec{\bar{P}}+\vec{\bar{P}}\cdot\vec{\bar{M}}$ because this term is zero as can be seen from the fact that (see \cite{Magpantay})
\begin{equation}\label{sha51}
\bar{M}_i=\epsilon_{ijk}\bar{X}_j\bar{P}_k.
\end{equation}
Assume the following ansatz for the Casimir invariant
\begin{equation}\label{sha52}
\bar{\textbf{C}}=\bar{P}_0^2-\vec{\bar{P}}\cdot\vec{\bar{P}}+a\vec{\bar{N}}\cdot\vec{\bar{N}}+b\vec{\bar{M}}\cdot\vec{\bar{M}}+c(\vec{\bar{N}}\cdot\vec{\bar{P}}+\vec{\bar{P}}\cdot\vec{\bar{N}})+d(\vec{\bar{M}}\cdot\vec{\bar{N}}+\vec{\bar{N}}\cdot\vec{\bar{M}}).
\end{equation}
Taking the commutator with $\bar{P}_0$, we get $a=0$, $c=-\bar{\kappa}$, $d=0$. Taking the commutator with $\bar{P}_i$, we get $b=-\bar{\kappa}^2$. With these values, we find that the commutators with $\bar{M}_i$ and $\bar{N}_i$ are trivially satisfied. Thus we find that the Casimir invariant of the dual kappa Poincare algebra in the dual bicrossproduct basis is
\begin{equation}\label{sha53}
\bar{\textbf{C}}=\bar{P}_0^2-\vec{\bar{P}}\cdot\vec{\bar{P}}-\bar{\kappa}^2\vec{\bar{M}}\cdot\vec{\bar{M}}+-\bar{\kappa}(\vec{\bar{N}}\cdot\vec{\bar{P}}+\vec{\bar{P}}\cdot\vec{\bar{N}}).
\end{equation}
In the Poincare limit, i.e., $\bar{\kappa}\rightarrow 0$, we find equation (53) reducing to the Poincare first Casimir invariant, thus $\bar{\textbf{C}}=m^2$ and the Casimir invariant for the dual kappa Poincare algebra seems to be correct.

	Equation (53) is one of the main results of this paper. And we will use this result to derive the speed of light by following the same prescription in Poincare and kappa Poincare algebras. First we derive an expression for $\bar{N}_i$ in terms of the dual kappa Poincare coordinates and momenta. Using equations (20b), (34a), (34b) of \cite{Magpantay}, we find
\begin{equation}\label{sha54}
\bar{N}_i=-\frac{1}{2\bar{\kappa}}\bar{P}_i+\bar{X}_i\bar{P}_0-\frac{\bar{\kappa}}{2}\bar{P}_i(\vec{\bar{X}}\cdot\vec{\bar{X}})+\frac{1}{2\bar{\kappa}}\bar{P}_i\exp{(-2\bar{\kappa}\bar{X}_0)}.
\end{equation}
This expression for $\bar{N}_i$ has operator ordering ambiguity in the second and third terms because of the non-commutativity of $\bar{P}_0$ with $\bar{X}_i$ (see equation (50e)) and the non-comutativity of $\bar{X}_i$ with $\bar{P}_j$. And if we do not operator order these two terms, we find that the Lorentz algebra relations given by equations (1b) and (1c) will not be satisfied. The operator-ordered $\bar{N}_i$ that is consistent with the dual kappa Poincare algebra given by equations (1a) to (1g) is
\begin{equation}\label{sha55}
\bar{N}_i=-\frac{1}{2\bar{\kappa}}\bar{P}_i+\frac{1}{2}(\bar{X}_i\bar{P}_0+\bar{P}_0\bar{X}_i)-\frac{\bar{\kappa}}{4}\left(\bar{P}_i(\vec{\bar{X}}\cdot\vec{\bar{X}})+(\vec{\bar{X}}\cdot\vec{\bar{X}})\bar{P}_i\right)+\frac{1}{2\bar{\kappa}}\bar{P}_i\exp{(-2\bar{\kappa}\bar{X}_0)}.
\end{equation}
These consistency calculations, which make use of the dual kappa Poincare phase space algebra (equation (50)), are straightforward but rather long and tedious.

	Using equations (51), (55) and (1g), we express the Casimir invariant as
\begin{equation}\label{sha56}
\begin{split}
\bar{\textbf{C}}&=\bar{P}_0^2-(\vec{\bar{P}}\cdot\vec{\bar{P}})\exp{(-2\bar{\kappa}\bar{X}_0)}-\bar{\kappa}[(\vec{\bar{P}}\cdot\vec{\bar{X}})\bar{P}_0+\bar{P}_i\bar{P}_0\bar{X}_i+3i\bar{P}_0]\\
&+\frac{\bar{\kappa}^2}{2}[2(\vec{\bar{P}}\cdot\vec{\bar{P}})(\vec{\bar{X}}\cdot\vec{\bar{X}})+\bar{P}_i(\vec{\bar{X}}\cdot\vec{\bar{X}})\bar{P}_i-2(\vec{\bar{X}}\cdot\vec{\bar{X}})(\vec{\bar{P}}\cdot\vec{\bar{P}})-2i(\vec{\bar{X}}\cdot\vec{\bar{P}})+2(\vec{\bar{X}}\cdot\vec{\bar{P}})^2]=m^2.\\
\end{split}
\end{equation}
From this expression, we find the speed of light in dual DSR
\begin{equation}\label{sha57}
\left(\frac{\partial{\bar{P}_0}}{\partial{\bar{P}}}\right)_{m\rightarrow 0}\approx\exp{(-\bar{\kappa}\bar{X}_0)}
\end{equation}
Thus, we find that the speed of light in the dual bicrossproduct basis of the dual kappa Poincare algebra is the coordinate velocity of light in the planar coordinates as given by $ds^2=0$ with
\begin{equation}\label{sha58}
ds^2=d\bar{X}_0^2-\exp{(2\bar{\kappa}\bar{X}_0)}d\vec{\bar{X}}\cdot d\vec{\bar{X}}.
\end{equation}
This result is expected because the spacetime of the dual kappa Poincare algebra in the dual bicrossproduct basis is the planar coordinates of de Sitter gravity.
\section{\label{sec:level4}Scalar Field Theory}
	In the Poincare case, the scalar field theory - the Klein-Gordon equation, is derived from the first Casimir invariant and a differential realization of the momentum operator, which in turn follows from the phase space algebra. In the kappa Poincare algebra, the field theory is rather involved because of the non-commuting spacetime. Expressed in terms of Minkowski spacetime, the scalar field theory of kappa Poincare is given by the non-local expression \cite{Kowalski-Glikman4}
\begin{equation}\label{sha59}
S=\int d^4x\left\{\frac{1}{2}(\partial{_\mu}\phi)^\ast\sqrt{1+\partial^2}(\partial{_\mu}\phi)+\frac{m^2}{2}\phi^\ast\sqrt{1+\partial^2}\phi\right\},
\end{equation}
where $\partial^2=\partial_0^2-\vec{\partial}\cdot\vec{\partial}$.

	In the case of dual kappa Poincare algebra, we do not expect a non-local field theory because space-time is commutative. Also, given the results in section II, we expect a local field theory in de Sitter gravity. This will be shown stating from the Casimir invariant given by equation (49). The phase space algebra given by equation (43) is realized by the following differential realization of the momentum operators
\begin{subequations}\label{sha60}
\begin{gather}
\bar{P}_0=i\frac{\partial}{\partial{\bar{X}_0}}-i\bar{k}\bar{X}_j\frac{\partial}{\partial{\bar{X}_j}},\label{first}\\
\bar{P}_i=-i\frac{\partial}{\partial{\bar{X}_j}}.
\end{gather}
\end{subequations}
Substituting these in equation (56), we find the scalar field equation
\begin{equation}\label{sha61}
\left[-\frac{\partial^2}{\partial{\bar{X}_0}^2}-3\bar{\kappa}\frac{\partial}{\partial{\bar{X}_0}}+\exp{(-2\bar{\kappa}\bar{X}_0)}\frac{\partial^2}{\partial{\bar{X}_i}^2}-m^2\right]\phi=0.
\end{equation}
This is precisely the Klein-Gordon equation in de Sitter space-time given by
\begin{equation}\label{sha62}
\left[\frac{1}{\sqrt{\left|g\right|}}\partial_\mu\left(\sqrt{\left|g\right|}g^{\mu\nu}\partial_\nu\right)-m^2\right]\phi=0,
\end{equation}
with the metric given by equation (58). Thus, the field theory derived from the Casimir invariant of the dual kappa Poincare algebra in the dual bicrossproduct basis is de Sitter scalar field theory in planar coordinates.
\section{\label{sec:level5}Minimum Momentum}
	Physics at the Planck length - strings, quantum gravity, DSR - require the modification of the uncertainty principle (UP) to what is now called generalized uncertainty principle (GUP) because the Planck length, which provides a natural limit to spatial measurements, will involve tremendous energy that will necessarily require gravity to be quantized. In the case of DSR, which is called a bottom-up approach to quantum gravity, the Planck length is observer-independent. In dual DSR or de Sitter relativity, the biggest length scale is the de Sitter length, $\textit{l}\approx 10^{26} m$. By de Broglie's argument, we can associate a wave of momentum $P_{min}=h/\textit{l}$. In the present universe, these 'photons' with $P_{min}$ must have been remnants of some primordial photons (the tail end of the microwave background radiation pattern) created during the Big Bang, which were 'stretched' so to speak as the universe expanded to its present size. Since the wavelength is the size of the universe, these photons provide a static (frequency of the order of $10^{-18} sec^{-1}$, very weak uniform background electromagnetic field that carry very little energy carry (will not create pairs of particles), thus, their presence is hardly noticed. But if the Planck length provides a natural limit to spatial measurement and results in the generalized uncertainty principle, the minimum momentum should also provide a natural limit to measurement of momentum of particles because having such a momentum will disturb particles and as we will show here will lead to a change in the uncertainty principle to what we will call dual generalized uncertainty principle (dGUP). And in analogy with DSR, this minimum momentum must be observer-independent. This easily follows from the de Broglie relation and the fact that in de Sitter relativity the de Sitter length is an invariant. 
	
	We now give two heuristic derivations of the dual generalized uncertainty principle. First, consider a one dimensional system. The momentum $p=mv$ involves measuring the position $x$ at two times, $t$ and $t+\delta t$, and letting $\delta t\rightarrow0$, i.e.,
\begin{equation}\label{sha63}
p=m\left\{\frac{x(t+\delta t)-x(t)}{\delta t}\right\}.
\end{equation}
Unfortunately, in quantum mechanics, the time resolution is limited by the uncertainty principle $\Delta E\delta t=\hbar$ and thus we really cannot let $\delta t\rightarrow0$. The limit to time resolution is set by the Planck time, $10^{-43} sec$, which means quantum gravity effects must be taken into account. We will assume that the time resolution is much greater than the Planck time so we can neglect quantum gravity effects. The uncertainty in momentum measurement is given by
\begin{equation}\label{sha64}
\Delta p=m\left\{\Delta v+\frac{1}{2}\Delta a\delta t+...\right\}.
\end{equation}
The first term is the usual uncertainty in momentum and is thus equal to $\frac{\hbar}{\Delta x}$. The second term is $\Delta F \delta t$, the uncertainty in the impulse on the particle while the measurement is being done and this must be due to the presence of 'stretched' out photon of minimum momentum.  But note that since the uncertainty in position of the particle is $\Delta x$, the imparted impulse from the 'stretched' out photon of wavelength \textit{l}must only be $P_{min}\dfrac{\Delta x}{\textit{l}}$ because the momentum density of the 'stretched' out photon is $\dfrac{P_{min}}{\textit{l}}$. Thus, we find   
\begin{equation}\label{sha65}
\Delta p=\frac{\hbar}{\Delta x}+\frac{P^2_{min}}{\hbar}\Delta x,
\end{equation}
where we neglected a factor of $\frac{1}{2}$. 
	
	Equation (65) can be derived in another way. Consider the three dimensional case. The expression given by equation (64) will be considered as the uncertainty along one direction, say, along the x axis. As in the first derivation, we will attribute the second term to the so called 'stretched out' photon, in particular to its electric field (magnetic force is generally much weaker because $B_0=\frac{E_0}{c}$), which means $\Delta F_x=q\Delta E_x$, where q is the charge of the particle and $\Delta \vec{E}$ is the uniform, weak background electric field. The energy carried by this field is $cP_{min}$ with energy density $\epsilon_0\Delta E^2$, where $\epsilon_0$ is the electric permittivity of the vacuum. Since the volume is $\frac{4}{3}\pi\textit{l}^3$, and the fact that a particle's velocity is typically $10^{-2}$ of the speed of light (fast but non-relativistic to neglect magnetic force), the second term of equation (64) becomes 
\begin{equation}\label{sha66}
2nd term=b\frac{P^2_{min}}{\hbar}\Delta x,
\end{equation}     
where the proportionality factor b given by $10^{2}\sqrt{\dfrac{\pi q^2}{4h\epsilon_0c}}$ is of order 1. This follows by putting in the values of $\hbar$, $\epsilon_0$, c and the fundamental electric charge.
  	
	Equation (65) is the dual version of the generalized uncertainty principle (GUP). The GUP (see \cite{Garay} for guide to the original literature and \cite{Adler} for a heuristic derivation) is given by
\begin{equation}\label{sha67}
\Delta x=\frac{\hbar}{\Delta p}+\frac{L^2_{pl}}{\hbar}\Delta p.
\end{equation}
The corrections in both (GUP) and (dGUP), the second terms of euqations (65) and (67), are both extremely small. The (GUP) correction arise from quantum gravity effects, which make space-time non-commutative. The (dGUP) correction arise from the curvature effects of the cosmological constant, which leads to non-commuting momentum. 

	GUP was shown to follow from a modified Heisenberg algebra, typically written as
\begin{equation}\label{sha68}
\left[x_i,p_j\right]=i\hbar\left\{\delta_{ij}+\alpha\delta_{ij}p^2+\beta p_ip_j\right\},
\end{equation}
where the corrections to the Heisenberg uncertainty principle, the $\alpha$ and $\beta$ terms, are quantum gravity corrections dependent on the Planck length. A number of derivations of the Heisenberg algebra had been presented in the literature \cite{Maggiore}, \cite{Kempf}, \cite{Cortes}. The dual version of the generalized uncertainty principle (dGUP) must also be derivable from a deformation of the Heisenberg algebra that takes into account space-time curvature. And indeed there is such a deformation and it was presented a number of years ago by Aldrovandi and collaborators \cite{Aldrovandi2}. Its form is similar to equation (68), only the corrections involve the coordinates and the constants are dependent on the de Sitter length (or cosmological constant or $\bar{\kappa}$).

	Before we end this section, we note that the derivation of (dGUP) makes use of the 'stretched' out photon which has minimum momentum. But in the evolution of the universe from the Big Bang to the present, both gravity and electromagnetism remained long-range forces while the other two (strong and weak) became very short-range forces. If there are photons that 'stretched' out as the universe expanded (the analogy is the sound of a guitar string as the player moves the finger to lower fret resulting in sound of lower tone), there must also be gravitons that did the same. Unfortunately, the heuristic arguments used in deriving the (dGUP) rely on energy and momentum densities, local quantities not well-defined in gravity. This is the reason that we ascribed the minimum momentum to the 'stretched' out photons. Besides, it is clear from the microwave background radiation pattern that very long wavelength EM wave exists while gravity waves has not been found yet. Still, there must be a proof of (dGUP) that should not depend whether the particle is electrically charged or not. The effect of curvature due to the cosmological constant on short-distance physics as expressed through (dGUP) must be derivable regardless of the technical problem of defining a local energy momentum tensor in gravity.         		
\section{\label{sec:level6}Conclusion}	    
	This paper discussed the physics of dual kappa Poincare algebra, which turned out to be de Sitter relativity as the discussions in Sections II to IV show. However, the results are derived using the methods of DSR. This paper contributed the following to the topic of de Sitter relativity: (1) showed another derivation of the de Sitter algebra by making use of the Heisenberg double contruction and the co-algebra and then followed by an isomorphic transformation, (2) showed that the different coordinate systems used in de Sitter space-time satisfy the general condition satisfied by dual kappa Poincare space-time, (3) derived the co-product rules for the conformal, natural and Beltrami coordinates of de Sitter relativity, and (4) introduced the concept of observer-independent mimimum momentum and gave a physical origin for such a momentum, (5) and derived heuristically the change in uncertainty principle, the dual (GUP), which represents the effect of curvature due to the cosmological constant on short-distance physics.
	
	This paper then presented de Sitter relativity as dual to DSR. The derivations to arrive at de Sitter physics followed the methods in DSR. Given the fact that de Sitter relativity came way ahead of DSR, actually we could alternatively say that DSR is dual to de Sitter relativity.    	
\begin{acknowledgments}
I acknowledge the support provided by the Creative Work and Research Grant of the University of the Philippines System and office space provided by the Natural Sciences Research Insitute of the University of the Philippines, Diliman Campus.
\end{acknowledgments}

\end{document}